\documentclass[twocolumn,showpacs,aps,amsmath,amssymb]{revtex4}
\begin{document}
\pagestyle{myheadings}
\title{\bf INTERACTION of 4-ROTATIONAL GAUGE FIELD with ORBITAL MOMENT, GRAVI-DIAMAGNETIC EFFECT 
and ORBIT EXPERIMENT "GRAVITY PROBE B"}
\author{O.V. Babourova}\email{baburova@orc.ru}
\author{B.N. Frolov}\email{frolovbn@orc.ru}
\affiliation{Moscow State Pedagogical University,Department of Physics for Nature Sciences,Krasnoprudnaya 14, Moscow 107140,
\\ Russian Federation}

\begin {abstract}
A direct interaction of the 4-rotational (Lorentzian) gauge field with the angular orbital momentum of an external field is considered. 
This interaction appears in a new Poincar\'{e} gauge theory of gravitation, in which tetrads are not true gauge fields, but represent to be 
some functions of the translational and 4-rotational gauge fields. The given interaction leads to a new effect: the existence of an electronic orbits 
precession under the action of an intensive external gravitational field  (gravi-diamagnetic effect), and also substantiates the existence of the direct interaction 
of the proper angular momentum of a gyroscope with the torsion field, which theoretically can be generated by the rotational angular momentum of the planet the Earth. The latter interaction can be detected by the experiment "Gravity Probe B" (GP--B) on a satellite orbit.  
\end {abstract}

\maketitle
Now an experiment "Gravity Probe B" (GP--B) on a satellite orbit is realized with the aim of studying features of movement of a rotating trial body (gyroscope) in a gravitational field of the Earth. The results of measurements now is analyzed. These results will be compared to the predictions of the general theory of relativity (GR) and its various generalizations. In \cite{http} one can read about GP--B: 
"The geodetic effect has previously been determined to ~1\% in complex studies of the Earth-Moon system around the Sun. GP--B aims to measure it to ~0.01\%". Therefore the given experiment gives hope of detection of deviations from GR. Recently in the paper \cite{Mao}, a possible difference of results of the mentioned above experiment from predictions of GR is offered to treat as the presence of spacetime torsion. Authors of this article assume that if the rotating body (such as a planet) can derive spacetime torsion, then the gyroscope could react to this torsion, and consequently an experiment with a gyroscope such as GPB could become the best means for definition of the most adequate non-standard gravitational theories with torsion. 

Necessity for a non-standard theory arises here because it is considered in the standard Poincar\'{e} gauge theory of gravitation \cite{Kib1}--\cite{HehlHKN} that the orbital angular momentum cannot be a source of a field in view of absence of translational invariance of the arising field equations. The given interdiction is stated in the well-known review \cite{HehlHKN}, where one can read: 
"...total angular momentum ... cannot couple to a gauge potential as a dynamical current in a Poincar\'{e} invariant manner for it is not a translation-invariant property of matter". 
Authors of the article \cite{Mao} name this opinion as the standard folklore in the modern theory of gravitational field. They believe that this opinion is not correct, in \cite{Mao}  one can read: "...a rotating body also generates torsion through its rotational angular momentum, and the torsion in turn affects the motion of spinning objects such as gyroscopes". But any substantiations of this non-standard thesis are absent in \cite{Mao}. 

Moreover, the derivation of a gyroscope motion under a torsion field in \cite{Mao} is based on the non-correct assumption that a gyroscope moves in a torsion field along autoparallel world-lines, along which the proper angular momentum of a gyroscope $S^\mu$ undergoes parallel transport by the full Riemann--Cartan connection $\nabla_\lambda$ (with torsion) along its trajectory:
\begin{equation}
u^\lambda \nabla_\lambda S^\mu = 0\,.  \label{eq:moveS}    
\end{equation}                                                                 

This statement is wrong because it has been stated long ago \cite{Math}, \cite{Pap} (see also \cite{Dix}--\cite{Tr}) that even in Riemann spacetime there exists an interaction between a proper angular momentum of a gyroscope and a Riemann curvature tensor (the Mathisson force), which deflects a gyroscope from autoparallels. In spaces with torsion (a Riemann--Cartan space $U_4$), the equations of motion of a test particle with mass $m$ and spin $S^\mu$ can be derived (in the exterior form language) in the form \cite{YadPhys2} ($c = 1$):
\begin{eqnarray}
 u\wedge D\pi_{\mu}&=& - (1/2) (e_{\mu}\rfloor {\cal R}^{\sigma\rho})\wedge S_{\sigma\rho}u \nonumber \\
&-& (e_{\mu}\rfloor {\cal T}^{\sigma})\wedge \pi_{\sigma}\, , \label{eq:part}\\
 m u\wedge DS^{\mu} &=& - u^{\mu} \sigma^{nu} \bigl ( (1/2) (e_{\nu}\rfloor {\cal R}^{\sigma\rho})\wedge S_{\sigma\rho}u \nonumber \\ 
&+& (e_{\nu}\rfloor {\cal T}^{\sigma})\wedge \pi_{\sigma} u \bigr )\, . \label{eq:BMT}
\end{eqnarray}
Here $D$ is a Riemann--Cartan exterior form covariant differential, $e_{\mu}$ -- a basis vector, $\rfloor$ -- an operator of contraction, ${\cal R}^{\sigma\rho}$ -- a curvature 2-form, ${\cal T}^{\sigma}$ -- a torsion 2-form, $u=\bar u\rfloor \eta$ -- a 4-velocity 3-form ($\bar u$ -- a velocity 4-vector, $\eta$ -- a volume 4-form), $\pi_{\mu}=mu_{\mu} - S_{\mu\nu}\bar{u}\rfloor D u^\nu$ -- a generalized momentum vector, $S_{\mu\nu}$ -- a spin tensor, $S^\mu$ -- the Pauli--Lubanski spin vector:
$S^\mu = (1/2)\eta^{\mu\nu\sigma\rho}S_{\nu\sigma}u_\rho$ ($\eta^{\mu\nu\sigma\rho}$ are totally antisymmetric components of a volume 4-form, that is a totally antisymmetric Levi--Chivita tensor). In Eqs. (\ref{eq:part}), (\ref{eq:BMT}) the terms with the curvature 2-form appear due to the Mattisson force, the terms with the torsion 2-form appear due to the additional "translational" force existing only in Riemann--Cartan spacetime. As we see, Eq. (\ref{eq:BMT}) is much more complicated then Eq. (\ref{eq:moveS}) of the paper \cite{Mao}.

The above-mentioned difficulties have been overcome in a new variant of Poincar'{e} gauge \cite{Fr:book}--\cite{FrIv} and in a recently developed Poincar'{e}-Weyl gauge  \cite{BFrZuk1}--\cite{BFrZuk3} theories of gravitation,  in which an angular momentum equally with a spin momentum arises as a source of a gravitational field (including a torsion field), not being violated translational invariance of field equations. The theory is constructed on the basis of Noether theorems, allowing to introduce gauge fields dynamically realizing conservation laws of an energy-momentum and a total rotational momentum  (the sum of orbital angular and spin momenta). In this approach tetrads are not true gauge fields, but represent to be some functions from the true gauge fields (4-rotational r-field  $A^m_{a}$ and 4-translational t-field $A_{a}^{k}$):
\begin{eqnarray*}
&& h^{\mu}\!_{a} = \stackrel{\circ}{h}\!^{\mu}\!_{k} Y_a^k\, , \;\; Y^k_a = A^k_a + A^m_a I_{m}{}^{k}\!_{l}\,x^l\, , \\
&& Z^a_k = (Y^{-1})^a_k \, ,  \;\; h^a\!_{\mu} = (h^{-1})^a\!_{\mu} = Z^a_k \stackrel{\circ}{h}\!^{k}\!_{\mu}\,, 
\end{eqnarray*}
where $x^k$ -- Cartezian coordinates in tangent Minkowskian space, $I_{m}{}^{k}\!_{l}$ --  generators of Lorentz group vector representation, $\stackrel{\circ}{h}\!^{k}\!_{\mu}$ -- 
subsidiary tetrads of a flat Minkowskian space.  

The gauge field equations have the forms (${\cal L}_0$ and ${\cal L}_{\psi}$ are Lagrangian densities of a gravitational field and an external field $\psi^{A}$, accordingly, $g$ -- a determinant of a metric tensor):  
\begin{eqnarray}
&&\frac{\delta{\cal L}_0}{\delta A_{a}^{m}} = -\frac{\partial {\cal L}_{\psi}} {\partial A^m_{a}}
= \sqrt{\mid g\mid}\,(\stackrel{\circ}{M}\!^a\!_{m} + S^a\!_{m} )\, , \label{eq:istAm}\\
&&\stackrel{\circ}{M}\!^a\!_{m} = I_m{}^c\!_b\, x^b \stackrel{(g)}{t}\!^a\!_c\,, \nonumber \\
&&\sqrt{\mid g\mid}\, S^a\!_{m} = \frac{\partial{\cal L}_{\psi}}{\partial D_a \psi^{A}}I_{m}\!^{A}\!_{B}\psi^{B}\, ,\nonumber \\
&&\frac{\delta{\cal L}_0}{\delta A^k_{a}} = - \frac{\partial {\cal L}_{\psi}}{\partial A^k_{a}}
= \sqrt{\mid g\mid}\,\stackrel{(g)}{t}\!^a\!_k\, ,\label{eq:istAk} \\
&& \sqrt{\mid g\mid} \stackrel{(g)}{t}\!^{a}\!_k = Z^a_l \,\left ({\cal L}_{\psi}\delta^l_k -
\frac{\partial{\cal L}_{\psi}}{P_l \psi^A }\, P_k \psi^A \right )\, . \nonumber 
\end{eqnarray}
Here $\stackrel {(g)}{t}\!^{a}\!_k$ is an energy-momentum tensor, $S^a\!_{m}$ is a spin momentum,
$\stackrel{\circ}{M}\!^a\!_{m}$ is an orbital angular momentum, $ P_k = - \stackrel{\circ}{h}\!^{\mu}\!_k \,\partial_\mu $, and $D_a \psi^A$ is the gauge derivative: 
\begin{equation}
D_a \psi^A = h^\mu\!_{a} \partial_\mu \psi^A - A^m_a I_{m}\!^{A}\!_{B} \psi^B\, . \label{eq:Dm}
\end{equation}                                                         

In \cite{BFrZuk2}, \cite{BFrZuk3} a case is considered, when an external field is a spinor field. Then the Lagrangian density of interaction of this external field with the Poincar\'{e} gauge field looks like 
\begin{equation}
{\cal L}_{\psi}= \sqrt{\mid g\mid}L_\psi\,,\;\; L_\psi = \psi_A \gamma D_a \psi^A - m\psi_A \psi^A\,.  \label{eq:Lagpsi}    
\end{equation}                                                                 
                                             
Let's consider the gauge field as a weak field in a Cartesian system of coordinates. For tetrad fields it means the representation $h^\mu\!_{a} = \delta^\mu\!_{a} + h^{\mu(1)}\!_{a}$, where  $h^{\mu(1)}\!_{a}$ (equally with $A^k_a$ and $A^m_a$) is an infinitesimal small value of the first order. Having substituted in the Lagrangian density (\ref{eq:Lagpsi}) the expression (\ref{eq:Dm}), we shall receive (equally with the standard terms in Lagrangian density describing, for example, a direct interaction of a gravitational field with the spin momentum of an external field) also additional terms, in particular, the following term:
\begin{equation}
L_{\psi M} = A^m_a\, I_m\!^k\!_l\,x^l \bar{\psi}_A \gamma^a h^\mu\!_k \partial_\mu \psi^A \,. \label{eq:IntAM}
\end{equation}  
This term  describes in the gravitational Lagrangian density a direct interaction of the 4-rotational gauge field with the angular orbital momentum of an external spinor field. As it has been already specified, the term of such kind is absent in standard GR.

Let's calculate this additional (in comparison with standard GR) interaction in the Lagrangian density. The calculation of the energy-momentum tensor of a spinor field results in the expression
\begin{equation}
t^a_k = Z^a_k L_\psi - \bar{\psi}_A \gamma^a \,h^\mu\!_k\,\partial_\mu \psi^A\,. \label{eq:tpsi}	
\end{equation}
Here any concrete expression for a matrix $Z^a_k$ is insignificant, because we have $L_\psi = 0$ by virtue of the spinor field equation. We shall transform (\ref{eq:IntAM}) with the help of (\ref{eq:tpsi}) and substitute in the result the expression for generators of Lorentz group vector representation:
\[
I_{ij}\!^a\!_b = \delta^a_i g_{jb} - \delta^a_j g_{ib} \,, \quad (m\rightarrow {i,j}\,,\;\; i<j)\,.
\]
Then we shall receive:
\begin{eqnarray}
&& L_{\psi M} = (1/2)A^{ij}_a M^a_{ij}\,, \label{eq:AM}\\
&& M^a_{ij} = x_i\,t^a\!_j - x_j\,t^a\!_i\,. \nonumber
\end{eqnarray}        

The theory predicts new effect of the direct interaction (\ref{eq:AM}) of the 4-rotation gauge field with the orbital angular momentum of external fields. The 4-rotational gauge field enters into the expression for spacetime torsion:
\[
T^{\lambda}\!_{\mu\nu} = 2h^\lambda\!_a ( \partial_{ [\mu} h^a\!_{\nu ]} + A^n_c I_n{}^a\!_b h^b\!_{[\mu} h^c\!_{\nu]}) \, .
\]
Therefore the given interaction will describe in the experiment "Gravity Probe B" the direct interaction of the full angular momentum of a gyroscope with the torsion field, which is in turn generated according to the equation (\ref{eq:istAm}) by the full angular momentum of the planet the Earth. 

The interaction described by the expression (\ref{eq:AM}) can be realized as an effect of precession of electronic orbits under action of an intensive external gravitational field (in particular, a torsion field). When this interaction is negligible, the plane of an orbit will occupy a constant position in space. At slow enough influence of the 4-rotational gauge field on the orbital momentum, a precession of the orbit will occur. This phenomenon will be perceived as change of diamagnetic properties of substance and can be designated as gravi-diamagnetic effect. By this effect the diamagnetic properties of patterns inside cosmic stations (where an inertial frame is realized) slightly differ from the diamagnetic properties of these patterns into Earth laboratories.  Discovery of this phenomenon may be an "experimentum crucis" for revealing the 4-rotational gauge field generated by rotating bodies. Also it is possible to assume that the gravi-diamagnetic effect can be used in principle for detecting gravitational waves and waves of torsion.

This effect is necessary to distinguish from an effect of interaction of elementary particle spin with a torsion field that is realized in standard Poincar\'{e} gauge theory of gravitation. In \cite{HehlHKN} it is possible to read: "A test particle in $U_4 $ theory, one which could sense torsion, is a particle with dynamical spin like the electron".  It is revealed long ago that the given interaction results to precession of electronic spins around of a pseudo-trace vector \cite{AdamTr}, \cite{KrPon} (when spacetime curvature is negligible). As a result there will be a change of spin orientation in paramagnetic medium in a direction of this vector. The given effect can be named gravi-paramagnetic effect and also can be used for detection torsion (and also curvature according (\ref{eq:BMT})).

\end{document}